\documentclass[usenatbib]{mnras}

\usepackage{ae}

\usepackage{graphicx}	
\usepackage{amsmath}	
\usepackage{amssymb}	

\title[Short title, max. 45 characters]{Hubble PanCET: An isothermal day-side atmosphere\\
 for the bloated gas-giant HAT-P-32Ab}
\author[N. Nikolov et al.]{N. Nikolov,$^{1}$\thanks{E-mail: nikolay@astro.ex.ac.uk (NN)},
D. K. Sing$^{1}$, 
J. Goyal$^{1}$, 
G. W. Henry$^{2}$,
H. R. Wakeford$^{1}$, 
T. M. Evans$^{1}$,\newauthor
M. L\'{o}pez-Morales$^{3}$, 
A. Garc\'{i}a Mu$\tilde{{\rm{n}}}$oz$^{4}$, 
L. Ben-Jaffel$^{5}$,
J. Sanz-Forcada$^{6}$, 
G. E. Ballester$^{7}$,\newauthor 
T. Kataria$^{8}$, 
J. K. Barstow$^{9}$,
V. Bourrier$^{10}$, 
L. A. Buchhave$^{11}$, 
O. Cohen$^{12}$,
D. Deming$^{13}$,\newauthor 
D. Ehrenreich$^{10}$,
H. Knutson$^{14}$,
P Lavvas$^{15}$, 
A. Lecavelier des Etangs$^{5}$, 
N. K. Lewis$^{16}$,\newauthor
A. M. Mandell$^{17}$
M. H. Williamson$^{2}$
\\
$^{1}$Physics and Astronomy, University of Exeter, EX4 4QL Exeter, UK\\
$^{2}$Tennessee State University, 3500 John A. Merritt Blvd., PO Box 9501, Nashville, TN 37209, USA\\ 
$^{3}$Harvard-Smithsonian Center for Astrophysics, 60 Garden Street, Cambridge, Massachusetts 02138, USA\\
$^{4}$Zentrum f\"ur Astronomie und Astrophysik, Technische Universit\"at Berlin, D-10623 Berlin, Germany\\
$^{5}$Sorbonne Universit\'es, UPMC Universit\'e Paris 6 and CNRS, UMR 7095, Institut d'Astrophysique de Paris, 98 bis boulevard Arago, F-75014 Paris, France\\
$^{6}$Centro de Astrobiolog\'ia (CSIC-INTA), ESAC Campus, Camino bajo del Castillo, E-28692 Villanueva de la Ca$\tilde{{\rm{n}}}$ada, Madrid, Spain \\
$^{7}$Lunar and Planetary Laboratory, University of Arizona, Tucson, Arizona 85721, USA\\
$^{8}$NASA Jet Propulsion Laboratory, 4800 Oak Grove Drive, Pasadena, California 91109, USA\\
$^{9}$Department of Physics and Astronomy, University College London, Gower Street, London WC1E 6BT, UK\\
$^{10}$Observatoire de l'Universit\'e de Gen\`eve, 51 chemin des Maillettes, 1290 Sauverny, Switzerland\\
$^{11}$Centre for Star and Planet Formation, Niels Bohr Institute and Natural History Museum, University of Copenhagen, DK-1350 Copenhagen, Denmark\\
$^{12}$Lowell Center for Space Science and Technology, University of Massachusetts, Lowell, Massachusetts 01854, USA\\
$^{13}$Department of Astronomy, University of Maryland, College Park, MD 20742 USA\\
$^{14}$Division of Geological and Planetary Sciences, California Institute of Technology, Pasadena, CA 91125 USA\\
$^{15}$Groupe de Spectroscopie Mol\'eculaire et Atmosph\'erique, Universit\'e de Reims, Champagne-Ardenne, CNRS UMR 7331, France\\
$^{16}$Space Telescope Science Institute, 3700 San Martin Drive, Baltimore, Maryland 21218, USA\\
$^{17}$NASA Goddard Space Flight Center, Greenbelt, Maryland 20771, USA\\
}

\date{Accepted XXX. Received YYY; in original form ZZZ}

\pubyear{2017}

\begin{document}
\label{firstpage}
\pagerange{\pageref{firstpage}--\pageref{lastpage}}
\maketitle

\begin{abstract}
We present a thermal emission spectrum of the bloated hot Jupiter HAT-P-32Ab from a single eclipse observation made in spatial scan mode with the Wide Field Camera 3 (WFC3) aboard the Hubble Space Telescope ({\it{HST}}). The spectrum covers the wavelength regime from 1.123 to $1.644\mu$m which is binned into 14 eclipse depths measured to an averaged precision of 104 parts-per million. The spectrum is unaffected by a dilution from the close M-dwarf companion HAT-P-32B, which was fully resolved. We complemented our spectrum with literature results and performed a comparative forward and retrieval analysis with the 1D radiative-convective {\tt{ATMO}} model. Assuming solar abundance of the planet atmosphere, we find that the measured spectrum can best be explained by the spectrum of a blackbody isothermal atmosphere with $T_{{\rm{p}}}=1995\pm17$\,K, but can equally-well be described by a spectrum with modest thermal inversion. The retrieved spectrum suggests emission from VO at the WFC3 wavelengths and no evidence of the $1.4\mu$m water feature. The emission models  with temperature profiles decreasing with height are rejected at a high confidence. An isothermal or inverted spectrum can imply a clear atmosphere with an absorber, a dusty cloud deck or a combination of both. We find that the planet can have continuum of values for the albedo and recirculation, ranging from high albedo and poor recirculation to low albedo and efficient recirculation. Optical spectroscopy of the planet's day-side or thermal emission phase curves can potentially resolve the current albedo with recirculation degeneracy.
\end{abstract}




\section{Introduction}



Transit, secondary eclipse (occultation) and phase curve observations 
provide an unprecedented access to the 
chemical composition, scattering and absorption from clouds and hazes, 
vertical thermal structure and recirculation 
of planetary atmospheres beyond the solar system \citep{winn10a, seager10, heng17, pont13}. 
A comparative transmission study of ten hot-Jupiter exoplanets showed 
that similar to solar system planets, where clouds and hazes are 
present on every planet with an atmosphere, exoplanets 
also have clouds and hazes in a continuum with a lack of temperature  dependance \citep{sing16}. 
Complementary to transit spectroscopy, which is sensitive to absorbing and scattering 
constituents (e.g. atomic sodium and potassium, water vapour, clouds and hazes) 
located in the upper atmosphere at the day-night terminator 
\citep{sing11b, sing16, nikolov14, nikolov15, nikolov16, evans16, gibson17},
observations of an exoplanetary occultation, i.e. when the planet
passes behind its host star, probe the deeper and hotter layers 
of the day-side hemisphere. 
Most of the secondary eclipse observations are made for short-period 
hot-Jupiter exoplanets (i.e. with high temperature and large radii 
to produce deep eclipses) and at near- and mid-infrared wavelengths 
(mainly with the {\it{Hubble}} and {\it{Spitzer Space Telescopes}}), 
where the planets are bright and the host stars are correspondingly faint \citep{madhu14}. 
Sensitive to the deeper layers of the atmosphere secondary eclipse observations 
can constrain the vertical temperature structure, chemistry and heat recirculation 
of exoplanets \citep{burrows07b, fortney10, knutson08a}. 
High-precision observations can distinguish an isothermal (blackbody), 
decreasing with height (non-inverted) or increasing with height (inverted) 
thermal profiles. In the first case, the emission spectrum of the planetary atmosphere 
would be featureless with 
a lack of absorbing or emitting features (e.g. H$_2$O, CO, CH$_4$, depending on the wavelength). 
An isothermal or blackbody spectrum indicates that either the temperature remains 
constant at each layer, or alternatively that the same pressure level (altitude)
is probed by the spectrum, i.e. the radiation 
originates from a cloud deck and the measured brightness temperature
is the temperature of the cloud near the altitude where the cloud becomes optically thick, i.e. at $\tau\sim1$.
If the planetary atmosphere contains layers 
where temperature decreases uniformly with altitude, the
resulting thermal spectrum is expected to exhibit absorption features \citep{stevenson14b}. On the 
contrary, if a hotter layer is located above a cooler 
region (stratosphere) the first will produce a thermal spectrum with the same features observed in emission \citep{madhu14, evans17}. 
Theoretical studies have predicted that the spectroscopically active gasses TiO and VO 
could capture the incident stellar radiation and consequently have been 
proposed as the responsible constituent for thermally inverted stratospheres \citep{hubeny03, fortney08}.  
Previous studies have demonstrated that the population of hot Jupiters 
are a rather heterogenous group with some of them showing spectra 
consistent with thermal inversion layer caused by TiO \citep{haynes15} or an 
unknown absorber and others have spectra consistent with 
no inversion layers \citep{fortney08, zhao14, line14, line16}. 
Using {\it{HST}}\,WFC3 \cite{evans17} recently reported compelling evidence of detection of a 
stratosphere in the very hot ($T_{\rm{p}}=2700\pm10$\,K) Jupiter WASP-121b, where the 
$1.4\mu$m water feature has been resolved and observed in emission.

In this paper we report measurements from a single {\it{HST}}~WFC3 secondary eclipse observation of 
HAT-P-32Ab. Our measured thermal emission spectrum is 
consistent with radiation from an isothermal blackbody, but can equally-well 
be described by a model with thermal inversion.
This result is part of the {\it{HST}} Panchromatic Comparative Exoplanet Treasury (PanCET) program,
which targets 20 exoplanets for the first large-scale simultaneous UltraViolet, Optical, Infrared (UVOIR) comparative 
study of exoplanets \citep{wakeford17, evans17}. A major aim of PanCET is to produce 
one of the first comparative studies of 
clouds and hazes in exoplanet atmospheres over a wide range of 
parameters such as temperature, metallicity, mass, and radius.
This paper is organized
as follows: in Section\,\ref{observations} we report the observations, reductions and light curve analysis;
the thermal spectrum is reported and discussed in Section\,\ref{discussion}; and 
Section\,\ref{conclusions} presents our conclusions.

\section{Observations and analysis} \label{observations}

\begin{figure}
\includegraphics[trim = 0 0 0 0, clip, width = 0.46\textwidth]{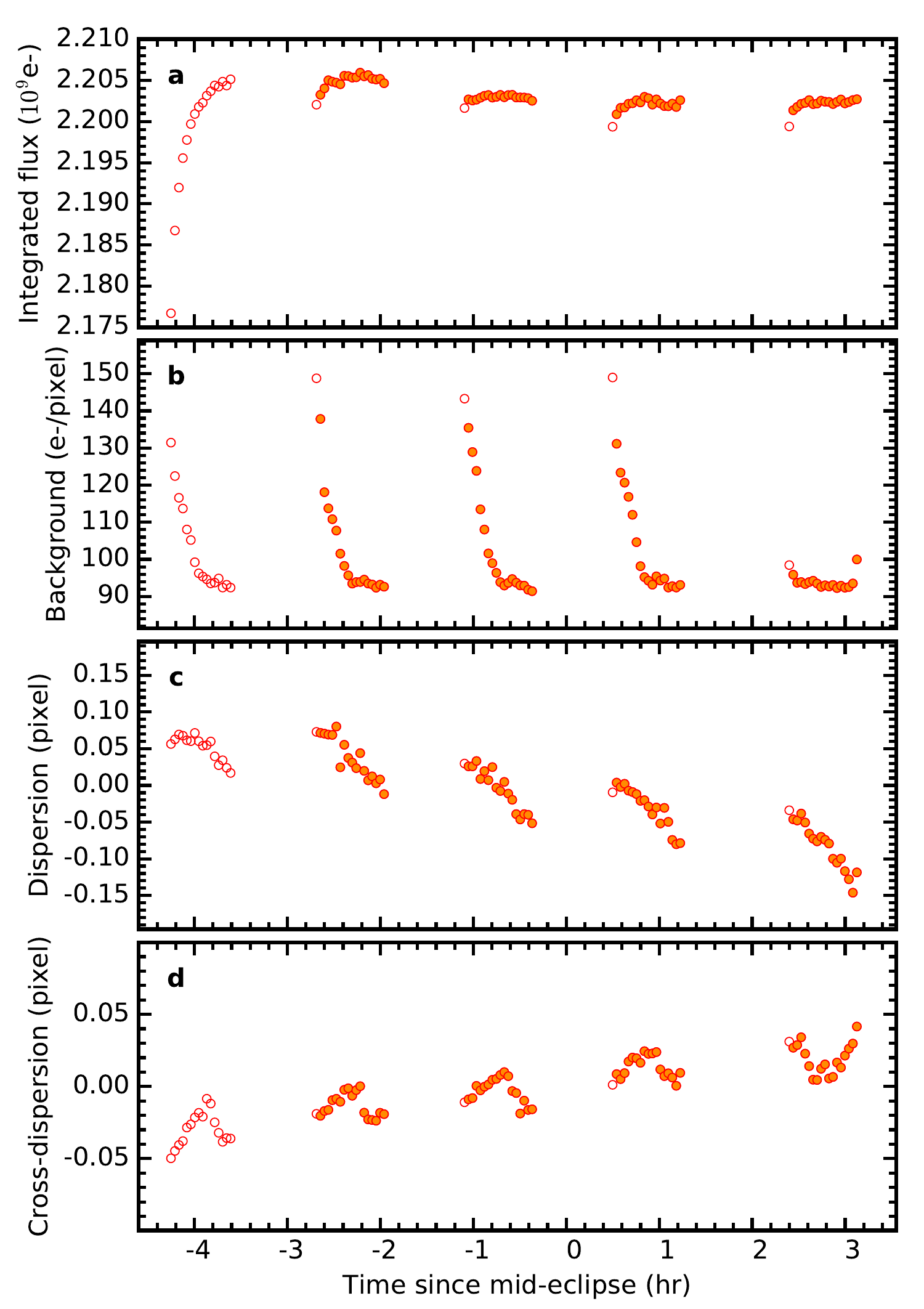}
\caption{White light curve and auxiliary variable time series for HAT-P-32A. Open symbols indicate
data that has been discarded and filled circles indicate the data that 
were retained for light curve analysis. {\bf{a:}} Raw flux band-integrated light curve in units of electrons. We 
discarded the first full {\it{HST}} orbit, because it exhibits particularly steep ramp as well as the first exposure of each
 subsequent orbit. {\bf{b:}} Measured background for each frame in units of electrons, using the last individual read 
 of each full scan. {\bf{c:}} Median-subtracted drift of the spectra along the dispersion axis in units of pixels, estimated
 by cross-correlating the spectra with a reference spectrum. {\bf{d:}} Same as the previous panel, but for 
 the cross-dispersion axis measured with flux-weighted mean.}
\label{fig:fig1}
\end{figure}

\subsection{The hot Jupiter HAT-P-32Ab}
The target of our study HAT-P-32Ab, is a standout highly inflated hot Jupiter ($M_{{\rm{p}}}=0.860\pm0.164\,M_{{\rm{Jup}}}$, 
$R_{{\rm{p}}}=1.789\pm0.025\,R_{{\rm{Jup}}}$ and an equilibrium temperature  $T_{{\rm{eq}}}=1786\pm26$\,K) 
on a circular orbit with period of $P=2.15$\,day around a moderately bright ($J=10.251\pm0.022$) late-F-type dwarf star in the northern constellation Andromeda \citep{hartman11, zhao14}. Observations of the Rossiter-McLaughlin effect showed
that the rotation axis of the star nearly lays in the plane of the planetary orbit $85^{\circ}\pm2^{\circ}$ 
implying a highly misaligned system \citep{albrecht12}.
Using adaptive optics \cite{adams13} detected a candidate M-dwarf companion (HAT-P-32B),
which was confirmed (at a separation of $2''.923\pm0''.004$ and position 
angle $110^{\circ}.64\pm0^{\circ}$, \citealt{zhao14}) and shown to be physically 
associated with HAT-P-32A \citep{knutson14}.
Combining broad-band ground-based {\it{Hale}} WIRC and {\it{Spitzer}} IRAC 3.6 and 4.5$\mu$m 
photometric observations of the secondary eclipse of  HAT-P-32Ab, \cite{zhao14} 
concluded that the thermal emission spectrum of the planet is equally well described
by a model assuming thermal inversion and a blackbody model with planet temperature $T_{{\rm{p}}}=2042\pm50$\,K. 
With one of the strongest transmission signals compared to similar hot Jupiters 
($\Delta\delta\approx2(H/R_{p})(R_{{\rm{p}}}/R_{\ast})^2\sim350$\,ppm for 1 pressure scale height, $H$, 
\citealt{winn10a}),\,~HAT-P-32Ab has been extensively followed up with optical transmission 
spectroscopy. While theory of irradiated gas giants (e.g. \citealt{fortney10}) predicted 
broad sodium, potassium and TiO/VO absorption features for clear atmospheres at visible wavelengths,
a number of optical studies reported a featureless flat transmission spectrum, suggesting a thick cloud deck at 
the day-night terminator of the planet \cite{gibson13a, Mallonn16a, Mallonn16b, Nortmann16}.

\subsection{{\it{HST}} WFC3 observations}
We observed a single secondary eclipse of HAT-P-32Ab with 
{\it{HST}} WFC3 on UT 2016 December 18 as part
of Program 14767 (PIs Sing and L$\rm{\acute{o}}$pez-Morales). 
Time series of spectra were collected with grism G141, which 
covers the wavelength range from $1.1\,\mu$m to $1.7\,\mu$m
at a resolution of $R=\lambda/\Delta\lambda=130$ and 
a dispersion of 4.7 nm pixel$^{-1}$.
The target was monitored for 7.4 hours during five consecutive orbits, 
covering the full eclipse, which lasted for 3.1 hours. 
The first, second and fifth {\it{HST}} orbits 
cover out of eclipse phases and the third and fourth orbits are in the eclipse.
Target acquisition was performed in imaging mode with an exposure 
time of 4.94s with the F139M filter. An inspection of the acquisition 
image showed the physically associated HAT-P-32B 
M-dwarf companion to be spatially resolved 
from the target. During each of the following 
spectroscopic observations, the telescope pointing was 
scanned along the cross-dispersion axis of the detector. 
This spreads the stellar flux across more pixels compared to a fixed 
pointing observation, allowing for a longer exposure 
times with higher duty cycle. We used forward scanning with a 
rate of 0.05 arcsec s$^{-1}$. To reduce
overheads we used a $256\times256$ subarray containing the target spectrum.
We used the SPARS10 sampling sequence with fourteen non-destructive reads 
per exposure (NSAMP=14) resulting in total integration times of 88.44\,s and
scans across 38 pixel rows of the cross-dispersion axis.
Typical count levels reached a maximum of $3.1 \times 10^4$ 
analog-to-digital units (ADU), i.e. well within the linear regime of the detector.
A total of 88 spectra were collected over the five {\it{HST}} orbits 
with 16 of them covering the first and 18 exposures obtained during 
each of the remaining four orbits. 

\subsection{Reductions and calibrations}
We started the analysis with the {\tt{ima}}\,2D spectra produced by the 
CalWFC3 pipeline (v3.1.6), which already had basic calibrations
including dark subtraction and flat-field correction. We extracted flux
for HAT-P-32A  from each exposure by taking the difference between successive 
non-destructive reads. For each read difference, 
we removed the background by taking the median flux in a box of pixels 
well away from the stellar spectra. Typical background levels integrated 
over the 88.44\,s exposures started between 
$\sim130-150$ electrons per pixel and decreased 
to $\sim90$ electrons per pixel over each HST orbit (see Figure\,\ref{fig:fig1}).
We then determined the flux-weighted centre of the HAT-P-32A scan and set to 
zero all pixel values located more than 14 pixels above and 
below along the cross-dispersion axis. 
Application of this top hat filter had the effect of masking the flux contributions 
from nearby contaminating stars including the companion HAT-P-32B located at $\sim23$ px
from the centre of the target. It had the additional advantage of eliminating many 
of the pixels affected by cosmic rays. The final reconstructed images were produced 
by adding together the read differences for each exposure.

Any remaining cosmic rays were identified by scanning each individual spectrum along the dispersion 
axis. For each scan we examined its difference with a median combined point-spread function (PSF) profile,
computed using the five preceding and five following consecutive columns. 
Before taking the difference we scaled the median profile to the scanned column.
Pixels deviating $>4\sigma$ were flagged as transient outliers and subsequently
replaced by their corresponding value from a nominal PSF profile, i.e. identical to the 
median combined PSF profile, but scaled to the column values, ignoring each of the flagged pixels as in \cite{nikolov14}.
Compared to other algorithms, e.g. temporal filtering, this cosmic ray identification approach 
has the advantage to be independent of spatial scan inhomogeneities, e.g.
read-out delays, which lead to extra flux accumulated typically in a few images of the time series.
We flagged between 0 and $\sim15$ pixels per reconstructed image. Our light curve analysis (see below)
was largely unaffected by whether or not we identified and corrected for cosmic ray events, although
for two of the images it reduced the scatter of the best-fit light curve residual.

We extracted target spectra using a fixed-size box by summing
the flux of all pixels. The box had dimensions of $176\times52$pixels,
and centred for each individual exposure. To 
identify the box position along the dispersion and cross-dispersion 
axis we took the flux-weighted mean of each 2D spectrum.
We found the target drifted on the dispersion and cross-dispersion axis by 
0.3 and 0.1 pixels, respectively (see Figure\,\ref{fig:fig1}). 

The wavelength solution was established by cross-correlating
each target spectrum against a Kurucz stellar spectrum model \citep{kurucz79}
with properties similar to the HAT-P-32 host star: T$_{{\rm{eff}}} =6207\pm88$\,K, 
$\log{g}=4.33\pm0.01$ and $\left[ {\rm{Fe}}/{\rm{H}}\right]=-0.04\pm0.08$, \citep{hartman11}.
\begin{figure}
\includegraphics[trim = 0 0 0 0, clip, width = 0.5\textwidth]{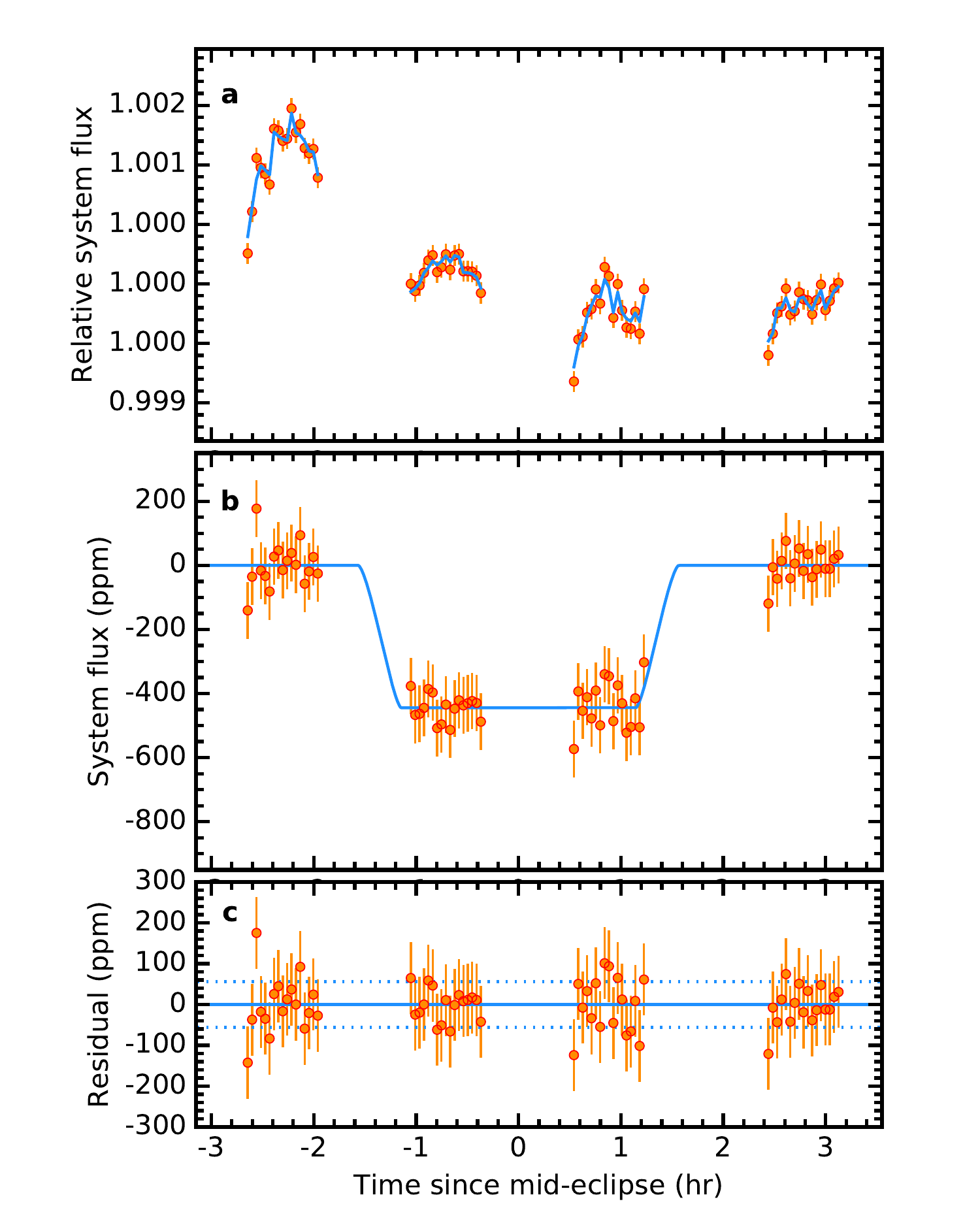}
\caption{White light curve for HAT-P-32Ab. {\bf{a:}} Raw flux, 
normalised to the median out-of transit baseline along with the photon noise 
uncertainties. The blue continuous lines indicate the best-fit GP model. The 
gaps in the data are a result of the target being occulted by the Earth during 
each {\it{HST}} orbit. {\bf{b:}} Normalised system flux and the best-fit
eclipse light curve after the best-fit GP systematics model has been removed. 
{\bf{c:}} Data minus best-fit model residuals with the fitted (rescaled photon noise) error-bars. The dotted lines
indicate the residual dispersion.}
\label{fig:fig2}
\end{figure}
\subsection{White light curve analysis}
We computed the WFC3/G141 band-integrated 'white' light curve by summing the flux of each stellar 
spectrum along the dispersion axis and across the wavelength 
range from 1.123 to 1.644 $\mu$m (Figure\,\ref{fig:fig2}). 
This wavelength range was chosen for the white and spectroscopic 
light curves to avoid the steep edges of the G141 grism response.
The white light curve exhibits quasi repeatable systematics which are considered to
originate from charge trapping in the detector {\citep{deming13, huitson13,zhou17}}.
We modeled the secondary eclipse and instrumental systematics
simultaneously by treating the data as a Gaussian process (GP) 
similar to the methodology of
\cite{gibson13a, gibson13b, gibson17, evans16, evans17}. The GP analysis in this study has been
performed with the Python GP library {\tt{george}} \citep{ambikasaran14, Foreman-Mackey15, corner}.
Under the GP assumption, the data likelihood is a multivariate normal distribution
with a mean function ${\boldsymbol{\mu}}$ describing the deterministic eclipse signal and a 
covariance matrix {\bf{K}} that accounts for stochastic correlations 
(i.e. poorly-constrained systematics) in the data:
\begin{equation}
    p( {\boldsymbol{f}} | {\boldsymbol{\theta}}, {\boldsymbol{\gamma}} )={\boldsymbol{\mathcal{N}}}( {\boldsymbol{\mu}}, {\mathbf{K}}),
	\label{eq:likelyhood}
\end{equation}
where $p$ is the probability density function, ${\boldsymbol{f}}$ is a vector containing the flux
measurements, ${\pmb{\theta}}$ is a vector containing the mean function parameters,
${\boldsymbol{\gamma}}$ is a function containing the covariance parameters, ${\boldsymbol{\mathcal{N}}}$ is a 
multivariate normal distribution. 
We defined the mean function ${\boldsymbol{\mu}}$ as follows:
\begin{equation}
    {\boldsymbol{\mu}(}t,\,{\boldsymbol{\hat{t}}};\,c_{\tiny{0}},\,c_{1},\,\delta, T_{{\rm{mid}}} ) = [c_0+c_1{\boldsymbol{\hat{t}}}] \,E(t;F_{{\rm{p}}}/F_{\ast}, T_{{\rm{mid}}}),
	\label{eq:meanf}
\end{equation}
Where $t$ is a vector of all central exposure time stamps in Julian Date (JD), ${\boldsymbol{\hat{t}}}$
is a vector containing all standardized times, i.e. with subtracted mean exposure time and dived by the 
standard deviation, $c_0$ and $c_1$ describe a linear baseline trend, 
$E$ is an analytical expression describing the secondary eclipse, $F_{{\rm{p}}}/F_{\ast}$ is the
planet-to-star flux ratio, $\delta=(F_{{\rm{p}}}/F_{\ast})(R_{{\rm{p}}}/R_{\ast})^2$ is the 
eclipse depth and $T_{{\rm{mid}}}$ is the eclipse central time. To obtain an analytical eclipse model $E$, 
we used a \cite{mandel02} transit model with limb darkening set to zero. We fixed the remaining 
system parameters to literature values listed in Table\,\ref{table:wlc_res} and fitted only for the eclipse depth and central time.
The covariance matrix is defined as 
\begin{equation}
    {\bf{K}} = \sigma_i^2 \delta_{ij}+k_{ij},
	\label{eq:cov}
\end{equation}
where $\sigma_i$ are the photon noise uncertainties, $\delta_{ij}$ is the Kronecker delta function and $k_{ij}$ is
a covariance function 
or kernel.
In our light curve fitting we assumed the white noise term was the 
same for all data points and allowed it to vary as a free parameter, $\sigma_w$. 
In this study we choose to use the Mat$\'e$rn $\nu=3/2$ kernel with {\it{HST}} orbital phase $\phi$, dispersion 
drift $x$ and cross-dispersion drift $y$ as input variables. As with the linear time term,
we also standardised the three input parameters ($\phi$, $x$ and $y$) prior to the fitting the light curve.
The covariance function then was defined as:
\begin{equation}
    k_{ij} = A^2\big(1+\sqrt{3}D_{ij}\big)\,{\rm{exp}}\big(-\sqrt{3}D_{ij}\big),
	\label{eq:cov_func}
\end{equation}
where $A$ is the characteristic correlation amplitude and 
\begin{equation}
	D_{ij} = \sqrt{ \frac{(\hat{\phi_i}-\hat{\phi_j})^2}{\tau^2_{\phi}} + \frac{(\hat{x_i}-\hat{x_j})^2}{\tau^2_{x}} + \frac{(\hat{y_i}-\hat{y_j})^2}{\tau^2_{y}}},
	\label{eq:covr}
\end{equation}
where $\tau_{\phi}$, $\tau_{x}$ and $\tau_{y}$ are the correlation length scales and the hatted variables are standardised. 
In the white light curve fitting we allowed parameters ${\boldsymbol{\theta}}=[c_0$, $c_1$, $F_{{\rm{p}}}/F_{\ast}$, $T_{{\rm{mid}}}]$ 
and ${\boldsymbol{\gamma}}=[A$, $\tau_{\phi}$, $\tau_{x}$, $\tau_{y}]$ to vary and fixed the  planet-to-star radius ratio 
$R_{{\rm{p}}}/R_{\ast}$, orbital inclination $i$ and orbital period $P$ to their literature values \citep{hartman11}. 
Our prior distribution had the form $p({\boldsymbol{\theta}}, {\boldsymbol{\gamma}} )=p(c_0)\,p(c_1)\,
p(F_{{\rm{p}}}/F_{\ast})\,p(T_{{\rm{mid}}})\,p(A)p(\tau_{\phi})\,p(\tau_{x})\,p(\tau_{y})$.
Uniform priors were adopted for $p(c_0)$, $p(c_1)$, $p(F_{{\rm{p}}}/F_{\ast})$, $p(T_{{\rm{mid}}})$. 
Log-uniform priors were adopted for $p(A)$, $p(\tau_{\phi})$, $p(\tau_{x})$, $p(\tau_{y})$.

Prior to fitting all light curves we discarded 
the data from the first full {\it{HST}} orbit, because of the particularly 
strong ramp systematic it exhibited. In addition, we also discarded the first data point of each of the 
subsequent four orbits as they had significantly lower flux compared to the 
rest of the data in the same orbit.

We used the Markov-Chain Monte Carlo (MCMC) Python software package {\tt{emcee}} \citep{Foreman13} to marginalise 
the posterior distribution $p( {\boldsymbol{\theta}}, {\boldsymbol{\gamma}} |  {\boldsymbol{f}} ) \propto p({\boldsymbol{f}} |  {\boldsymbol{\theta}}, {\boldsymbol{\gamma}} )p({\boldsymbol{\theta}}, {\boldsymbol{\gamma}})$. We initialised
three groups of 150 walkers close to the maximum likelihood solution, which was located 
using the Levenberg--Marquardt least-squares 
algorithm as implemented in the {\small{MPFIT\footnote{http://www.physics.wisc.edu/craigm/idl/fitting.html}}}. 
Groups one and two were run for 350 samples
and the third group had 2500 samples. Before running for the second group we 
re-sampled the positions of the walkers in a narrow space around the position of 
the best walker from the first run. 
This extra re-sampling step was useful because otherwise some of the walkers 
can start in a low likelihood area of parameter space and would require more 
computational time to converge. An eclipse model computed using the marginalized
 posterior distributions for the depth and central time is shown in 
 Figure\,\ref{fig:fig2} and the parameter values are reported 
in Table\,\ref{table:wlc_res}. We find the eclipse central time to be constrained
to only $\sim4.3$min. Such high uncertainty is explained with the insufficient phase 
coverage during the ingress and egress portions of the eclipse light curve.  
The best-fit model residuals from this analysis were found to be
within $2\%$ of the theoretical the photon noise expectation. The posterior distributions
for all fitting parameters of the white light curve are shown in Appendix\,\ref{appendix}. 
\begin{table}
	\centering
	\caption{System parameters}
	\begin{tabular}{lc}
		\hline
		Parameter     &    Value   \\
		\hline
		$P$\,(day) & 2.150009, fixed  \\
		$e$ & 0, fixed   \\
		$\omega$ ($^{\circ}$) & 0, fixed  \\
		$i$ ($^{\circ}$) & $88.90$, fixed  \\
		$a/R_{\ast}$ & $6.05$, fixed  \\
		$R_{{\rm{p}}}/R_{\ast}$ & $0.1508$, fixed   \\
		T$_{\rm{mid}}$ (JD) & $2457741.1194^{+0.0031}_{-0.0029}$  \\
		$(F_{{\rm{p}}}/F_{\ast})(R_{{\rm{p}}}/R_{\ast})^2$\,(ppm) & $445^{+72}_{-73}$  \\
		$c_0$  & $1.0003^{+0.00015}_{-0.00019}$    \\
		$c_1$  &  $(-4.5\pm6.9)\times10^{-5}$  \\
		A\,(ppm) &  $368^{+162}_{-93}$   \\
		$\ln \tau_{\phi}$  & $-0.32^{+1.04}_{-0.95}$    \\
		$\ln \tau_{x}$  &  $-0.3^{+1.1}_{-0.9}$  \\
		$\ln \tau_{y}$  &  $2.5^{+1.7}_{-1.5}$   \\
		$\sigma_{{\rm{w}}}$\,(ppm)  &  $57^{+39}_{-37}$   \\
		\hline
	\end{tabular}
	\label{table:wlc_res}
\end{table}

\subsection{Spectroscopic light curve analysis}
We produced 14 spectroscopic light curves across the 
wavelength range from 1.123 to 1.644 $\mu$m by summing the
flux of the stellar spectra in bins with widths of 8 pixels each (equivalent to $0.037\mu$m).
We removed wavelength-independent systematics 
using a common-mode correction. 
This is a powerful technique used in a number of 
previous works from both
space and the ground \citep{sing12, deming13, gibson13a, gibson13b, huitson13, nikolov15, nikolov16}.
To compute the common-mode we divided the raw white light curve
to an eclipse model. We computed the eclipse model using the central time $T_{{\rm{mid}}}$ and 
occultation depth from the marginalized posterior distributions of the white light curve fit 
and the adopted orbital period, inclination and normalised semimajor axis from Table\,\ref{table:wlc_res}.
Prior to fitting, we corrected the raw spectroscopic light curves by dividing each of them to the 
common-mode light curve (see Figure\,\ref{fig:fig3}).
The common-mode technique relies on the similarities 
of time dependent systematics, which can be characterised by the light curves 
themselves and removed individually for each spectral wavelength bin. Empirically 
determining and removing systematics has an advantage over a parameterised 
method, as higher order frequencies are naturally subtracted. 

\begin{figure*}
\includegraphics[trim = 0 0 0 0, clip, width = 0.99\textwidth]{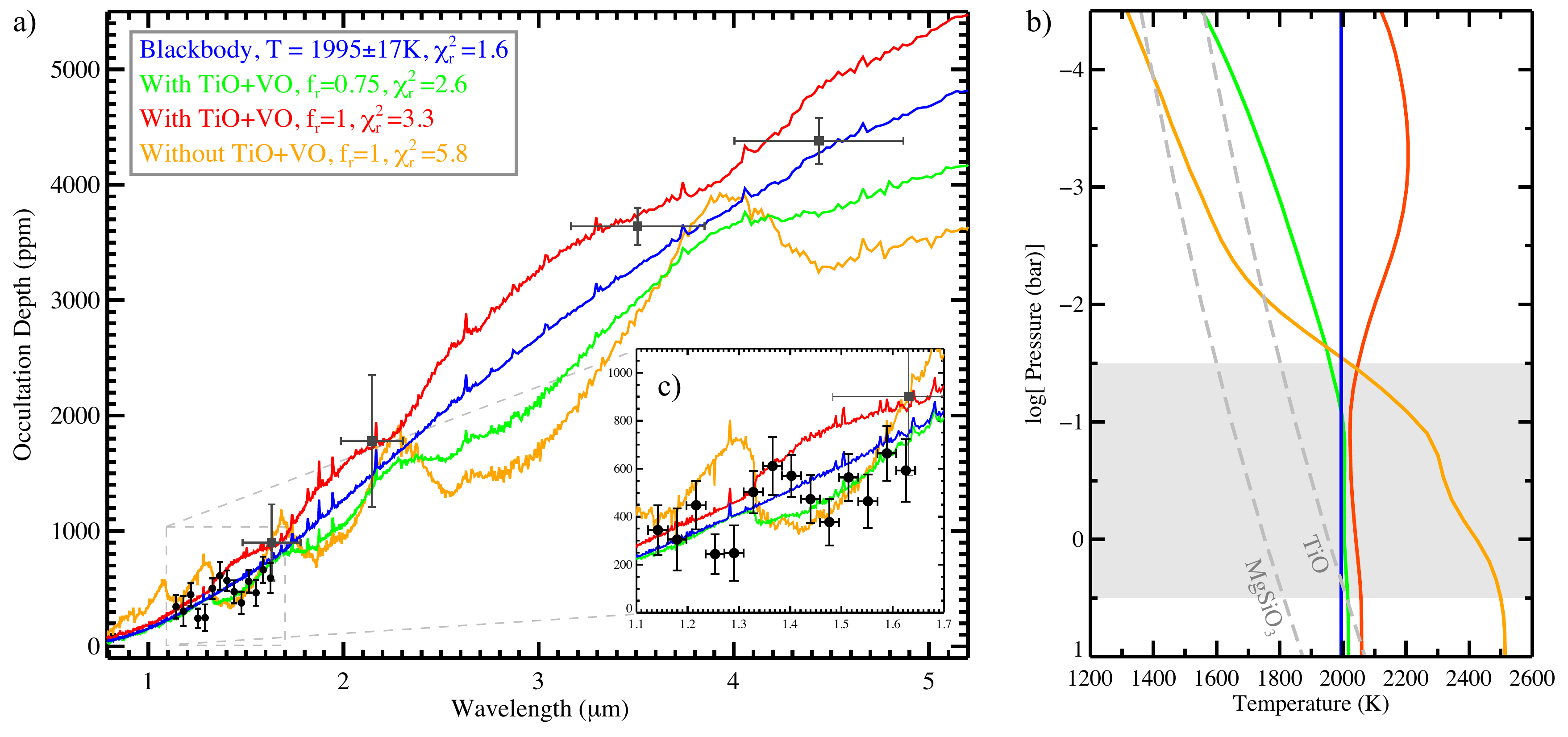}
\caption{Emission spectrum of HAT-P-32Ab, forward models and assumed pressure-temperature profiles;
a: Combined WFC3 emission spectrum of HAT-P-32Ab (black dots) with 
the dilution-corrected {\it{Hale}} WIRC $H$ and $K_S$ and {\it{Spitzer}}\,IRAC 3.6 and 4.5$\mu$m 
eclipse observations (grey dots) detailed in \citealt{zhao14}. The continuous lines indicate 
atmospheric models fit to the spectrum. The top two best-match models assume an 
isothermal and thermally-inverted pressure-temperature profiles. Models assuming decreasing 
temperature with an increasing altitude are excluded at high confidence; 
b:Zoom around the the WFC3 spectrum with models; and 
c: Pressure-temperature profiles assumed in the calculation of the emission models. 
The shaded region indicates the pressures probed by our WFC3 observations and the 
dashed lines indicate condensation curves for atmospheric constituents CaSiO$_3$ and TiO.}
\label{fig:fig4}
\end{figure*}

We performed fits to the common-mode corrected spectroscopic light curves adopting the same data likelihood  
and uniform prior distributions as for the white light curve analysis. However, we allowed
only the eclipse depth to vary and fixed the central time to its white light value. 
The results from our analysis are reported in Table~\ref{table:transpec} and the eclipse 
models, calculated using the parameter values from the corresponding marginalized distributions
are shown in Figure\,\ref{fig:fig3}. For all of the 14 bands we found a median scatter of 104 parts per million close
to the photon noise. The measured wavelength-dependent secondary eclipse (occultation) depths, which
comprise our thermal emission spectrum of HAT-P-32b 
are plotted in Figure\,\ref{fig:fig4}. 

\subsection{Ground-Based Photometry of HAT-P-32A}
Stellar activity can complicate the interpretation of transmission and emission spectra 
but can be corrected for if complimentary ground-based photometry is available 
\citep[e.g., HD~189733;][]{sing11b}. Therefore, we scheduled nightly
photometry of HAT-P-32A on the Tennessee State University Celestron 14-inch 
(C14) automated imaging telescope (AIT) at Fairborn Observatory 
\citep[see, e.g.,][]{henry99,eaton03}.  We have acquired a total of 223 
nightly observations during the 2014-15, 2015-16, and 2016-17 observing 
seasons.  The observations were made with a Cousins R filter and an 
SBIG STL-1001E CCD camera. Differential magnitudes were computed with respect 
to the mean brightness of 10 of the most constant comparison stars in the same 
field.  Further details of our data acquisition, reduction procedures, and 
analysis of the data can be found in \citet{sing15}, which describe a 
similar analysis of the planetary-host star WASP-31.

Our photometric observations are summarized in Table~\ref{table:activity}. Standard deviations 
of single nightly observations of HAT-P-32A with respect to their 
corresponding seasonal mean differential magnitudes are given in column~4; 
all are very close to 0.003 mag. This is the typical level of photometric 
precision achieved for a single observation with the AIT on good nights; 
see, e.g., Table~1 in both \citet{kreidberg15} and \citet{sing15}.  We 
performed periodogram analyses on each season and find no evidence for any
periodicity between 1 and 100 days.  In addition, the three seasonal means 
given in column~5 of Table~\ref{table:activity} agree to within a standard deviation of only 
0.000079~mag.  Therefore, we conclude that HAT-P-32A is constant, with the
exception of transits, on both nightly and yearly timescales to the limit of 
our photometric precision.


\begin{table} 
\caption{Summary of AIT photometric observations of HAT-P-32A}
\label{table:activity}
\begin{centering}
\renewcommand{\footnoterule}{}  
\begin{tabular}{ccccc}
\hline
 Observing &          &                   Date Range& Sigma & Seasonal Mean \\
Season      & $N_{{\rm{obs}}}$ &HJD$-$2400000& (mag) & (mag)\\
\hline
 2014-15  &  82 & 56943--57114 & 0.00302 & $-0.47210\pm0.00033$  \\
 2015-16  &  85 & 57293--57472 & 0.00301 & $-0.47185\pm0.00033$  \\
 2016-17  &  56 & 57705--57843 & 0.00283 & $-0.47063\pm0.00038$  \\
\hline
\end{tabular}
\end{centering}
\end{table}


\begin{table}
	\centering
	\caption{Thermal eclipse spectrum of HAT-P-32b}
	\label{table:transpec}
	\begin{tabular}{cc} 
		\hline
		Wavelength ($\mu$m)     &    Eclipse depth\,(ppm)   \\
		\hline
		$1.123-1.161$ & $344^{+100}_{-103}$ \\
		$1.161-1.198$ & $305^{+127}_{-125}$ \\
		$1.198-1.235$ & $448^{+95}_{-103}$ \\
		$1.235-1.272$ & $244^{+81}_{-83}$ \\
		$1.272-1.309$ & $248^{+110}_{-115}$ \\
		$1.309-1.347$ & $502^{+87}_{-87}$ \\
		$1.347-1.384$ & $611^{+118}_{-119}$ \\
		$1.384-1.421$ & $570^{+86}_{-87}$ \\
		$1.421-1.458$ & $473^{+101}_{-98}$ \\
		$1.458-1.495$ & $377^{+96}_{-94}$ \\
		$1.495-1.533$ & $564^{+95}_{-97}$ \\
		$1.533-1.570$ & $464^{+109}_{-113}$ \\
		$1.570-1.607$ & $664^{+113}_{-111}$ \\
		$1.607-1.644$ & $592^{+134}_{-126}$ \\
		\hline
	\end{tabular}
\end{table}

\section{Discussion}\label{discussion}
\subsection{Thermal spectrum}
We combined the WFC3 thermal emission spectrum with the dilution-corrected 
{\it{Hale}} WIRC $H$ and $K_S$
and {\it{Spitzer}} IRAC 3.6 and $4.5\mu$m eclipse observations detailed in \cite{zhao14}. 
To further reduce a potential offset in the eclipse depths 
of the {\it{Hale}}-{\it{Spitzer}} photometry and the {\it{HST}} spectrum
we adopted the system parameters of \cite{zhao14} 
when fitting the WFC3 light curves (see Table\,\ref{table:wlc_res}). 

\subsection{Blackbody and forward atmospheric models}\label{bbody}
We first fit a blackbody 
model to determine the
temperature of the planet. For this we used a
BT-Settl stellar model atmosphere \citep{allard12} with 
the closest match to the effective temperature,
surface gravity and metallicity of HAT-P-32A from 
\cite{hartman11}. 
We found that the best-fit model for the planet, 
with a $\chi^2=27$ for $17$ degrees of freedom, 
and BIC = 30.1, corresponds to a $T_{{\rm{p}}}=1995\pm17$\,K . 
This models correspond to the blue curve in Figure\,\ref{fig:fig4}
Our planet temperature 
measurement is consistent ($\sim0.9\sigma$) with 
the temperature reported in \cite{zhao14}, who found 
$T_{{\rm{p}}}=2042\pm50$\,K.

 

We then performed forward modelling of the 
measured thermal emission spectrum using the 1D atmosphere {\tt{ATMO}} code 
\citep{amundsen14, amundsen17, Tremblin15, Tremblin16, Drummond16, goyal17}. {\tt{ATMO}} computes the 1D 
pressure-temperature (P-T) atmospheric structure in plane-parallel geometry and  
also produces forward models assuming radiative, convective and chemical equilibrium.
The code includes isotropic multi-gas Rayleigh scattering, H$_2$-H$_2$ and H$_2$-He 
collision-induced absorption as well as opacities for all major chemical species 
taken from the most up-to-date high-temperature sources, including: H$_2$O, CO$_2$, CO,
CH$_4$, NH$_3$, Na, K, Li, Rb and Cs, TiO, VO, and FeH \citep{amundsen14, goyal17}. 
It uses the correlated-$k$ approximation with the random
overlap method to compute the total gaseous mixture opacity, which has been shown to agree well with a full
line-by-line treatment \citep{amundsen14}. 

We computed P-T profiles using 50 vertical model levels with minimum and maximum optical 
depth of $10^{-5}$ and $2\times10^5$ at $1\mu$m respectively. The P-T profiles were computed 
using 32 correlated-k bands while spectra were computed using 5000 correlated-$k$ bands 
equally spaced in wavenumber between 1 and $5\times10^4$\,cm$^{-1}$. Equilibrium chemistry 
calculations include condensation with rainout while computing P-T profiles. 

We computed {\tt{ATMO}} emission models,
assuming P-T profiles of a decreasing (non-inverted),  
and an increasing (inverted) temperature with altitude.
The models corresponding to each profile were computed assuming 
0.25, 0.5, 0.75 and 1 for the recirculation factors (f$_{\rm{r}}$), 
which govern the distribution of input stellar energy in the 
planet's atmosphere. A factor of unity implies no redistribution, while
factor of a half implies efficient redistribution \citep{fortney07}. 
In the calculation of the recirculation factor we assumed an angle $\mu=60^{\circ}.$

To compare models with observations we 
averaged the model spectra within the wavelength bins of the observed spectrum 
and computed the corresponding $\chi^2$. The results are shown in 
Figure\,\ref{fig:fig4}. From all the {\tt{ATMO}} forward models, 
the data is best fit by a model with TiO/VO and a weak thermal inversion. 
However, that model, with reduced chi-square of $\chi_{\rm{r}}^2=2.6$ and BIC=48.9, 
is a worse fit than the blackbody emission model described above
($\chi_{\rm{r}}^2=1.6$, BIC=30.1).  
This analysis also reveals that  {\tt{ATMO}} models without 
thermal inversions (the orange and green lines in Figure\,\ref{fig:fig4}), 
are unlikely to 
explain the observed spectrum. 
Finally, the planets' emission spectrum 
shows no evidence of water, based on these fits.




\subsection{Retrieval models}\label{retr}
Besides computing forward models, 
{\tt{ATMO}} can also be utilized as a retrieval tool to compute both emission and 
transmission spectra from an input P-T profile and arbitrary chemical abundances 
(see Section\,\ref{retr} and \citealt{wakeford17, evans17}).
Several of the WFC3 measurements, which have the highest precision 
from all data points, deviate at the $\sim2\sigma$ from the predicted 
absorption or emission water features. 
To further interpret the observed spectrum and constrain 
the planet P-T profile we turn to retrieval \citep{madhusudhan09, line13} analysis using {\tt{ATMO}}. 

Our retrieval approach follows the methodology detailed in \cite{evans17} and \cite{wakeford17}.
In summary, we fitted for the abundances of the molecules expected to add significant opacity in the wavelength region of our observations. In particular, we included H$_2$O, CO, CO$_2$, CH$_4$, NH$_3$, HCN, TiO and VO, assuming that those gasses are well-mixed vertically in the atmosphere. We assumed 50 pressure levels evenly spaced 
in log pressure between $10^{-8}$ and 500 bar
in the model planetary atmosphere. For the P-T profile, we adopted the 
1D analytic formulation of \cite{guillot10}, 
which assumes radiative equilibrium and is flexible enough to 
describe atmospheres with and without thermal inversion (stratosphere).
When using the parameterised P-T profile, we fitted the data assuming either one or two visible 
channels. 
With this assumption we had three to five parameters for the P-T profile: the Planck mean thermal
infrared opacity, $\kappa_{{\rm{IR}}}$; the ratios of the optical to infrared opacities in the two channels, 
$\gamma_1$ and $\gamma_2$; a partition of the flux between the two optical 
channels, $\alpha$; and an irradiation efficiency factor, $\beta$. 
We assumed a value of  $1.789R_{\rm{J}}$ for the planetary radius, corresponding to the 
lowest-altitude measured in transit observations at optical wavelengths. To model the input 
flux from the HAT-P-32A host star we assumed the same 
Phoenix stellar model as described in Section~\ref{bbody}. 

For all free parameters in our model we adopted uniform priors  with 
the following ranges: $10^{-5}$ to $10^{-0.5}$ for $\kappa_{{\rm{IR}}}$; 
$10^{-4}$ to $10^{1.5}$ for $\gamma_1$ and  $\gamma_2$; $0$ to $1$ 
for $\alpha$; and $0$ to $2$ for $\beta$. For the mixing ratios of chemical 
species other than H and He we adopted uniform priors between 
$10^{-12}$ and $0.05$. Our assumption on the metal abundances is motivated 
by the fact that HAT-P-32Ab is known to be a gas giant \citep{hartman11}.
Our retrieval analysis proceeded by first identifying the minimum $\chi^2$ 
solution using nonlinear least-squares optimization and then marginalizing 
over the posterior distribution using differential-evolution Markov chain Monte 
Carlo \citep{Eastman12}. A total of $10$ chains for $30,000$ steps each,
were ran until the Gelman-Rubin statistic for each free parameter was within $1\%$ of unity, showing
that the chains were well-mixed and had reached a steady state. 
 We discarded a burn-in phase from all chains corresponding to the step at which all chains had
found a $\chi^2$ below the median $\chi^2$ value of the chain \citep{Eastman12}. 
Finally, we combined the remaining samples into a single chain, forming our posterior distributions.

\begin{figure*}
\includegraphics[trim = 0 0 0 0, clip, width = 0.9\textwidth]{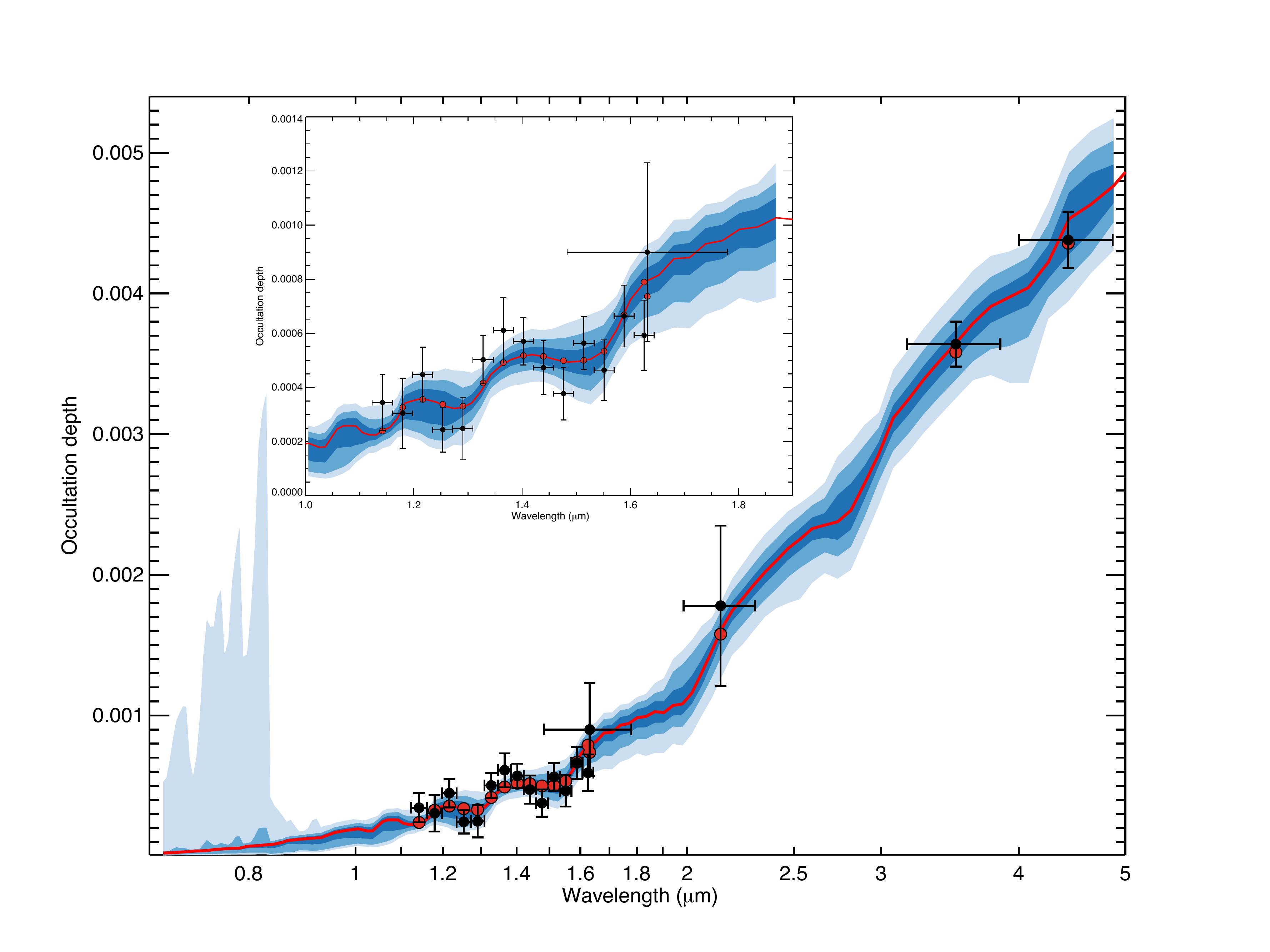}
\caption{Emission spectrum of HAT-P-32Ab (dots with $1\sigma$ uncertainties) with model 
emission spectra (lines) binned to the resolution of the data (red dots), obtained during the 
retrieval analysis. The continuous red line shows the best-fit retrieved model along with 1-, 2- and 3-$\sigma$ confidence 
levels. A zoom around the WFC3 is shown in the top left corner. }
\label{fig:fig9}
\end{figure*}

\begin{figure}
\includegraphics[trim = 0 0 0 0, clip, width = 0.47\textwidth]{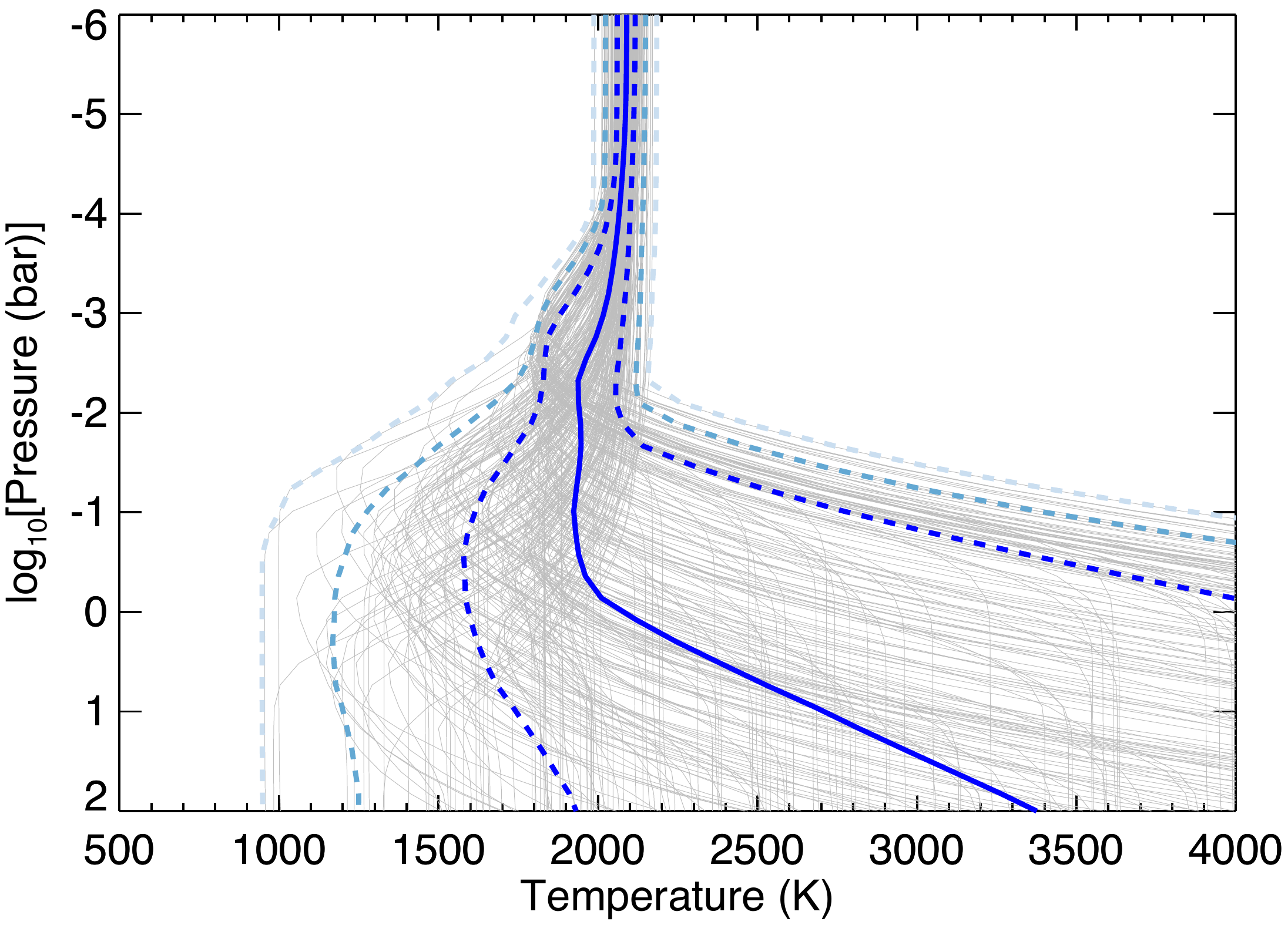}
\caption{Pressure-Temperature profiles for HAT-P-32Ab.
A subset of P-T profiles sampled during the MCMC retrieval analysis are indicated with the grey continuous lines. The median and 1-, 2- and 3-$\sigma$ confidence regions are indicated with the blue continuous and dashed lines, respectively.}
\label{fig:fig12}
\end{figure}

We found the retrieved abundances of all molecules to be poorly constrained, except for  CH$_4$ and VO, which are constrained between 0.5 and 2 orders of magnitude, respectively. The best-fit model to the data gave $\chi^2=11.16$ for 6 degrees of freedom and BIC=45.8. While CH$_4$ is preferred by the retrieval model, the lower 3-$\sigma$ bound on the CH$_4$ vertical mixing ratio is $1.1\times10^{-7}$.  This value is close to the solar-abundance chemical equilibrium value of CH$_4$ at 2000\,K and 1-bar \citep{sharp07}, so we do not have solid evidence for enhanced CH$_4$.
Like in the forward-modeling described in Section\,\ref{bbody}, retrieval models also give an isothermal P-T profile as the most favorable scenario, with some inverted profiles included within the 3-$\sigma$ confidence region. As in Section\,\ref{bbody}, retrieval models also disfavor non-inverted P-T profiles. We report our retrieval results in Figures~\ref{fig:fig9} and \ref{fig:fig12} along with the MCMC posterior distributions, shown in Figure\,\ref{fig:fig10}. 

\subsection{Constraining the albedo and recirculation}
\begin{figure*}
\includegraphics[trim = 0 0 0 0, clip, width = 0.8\textwidth]{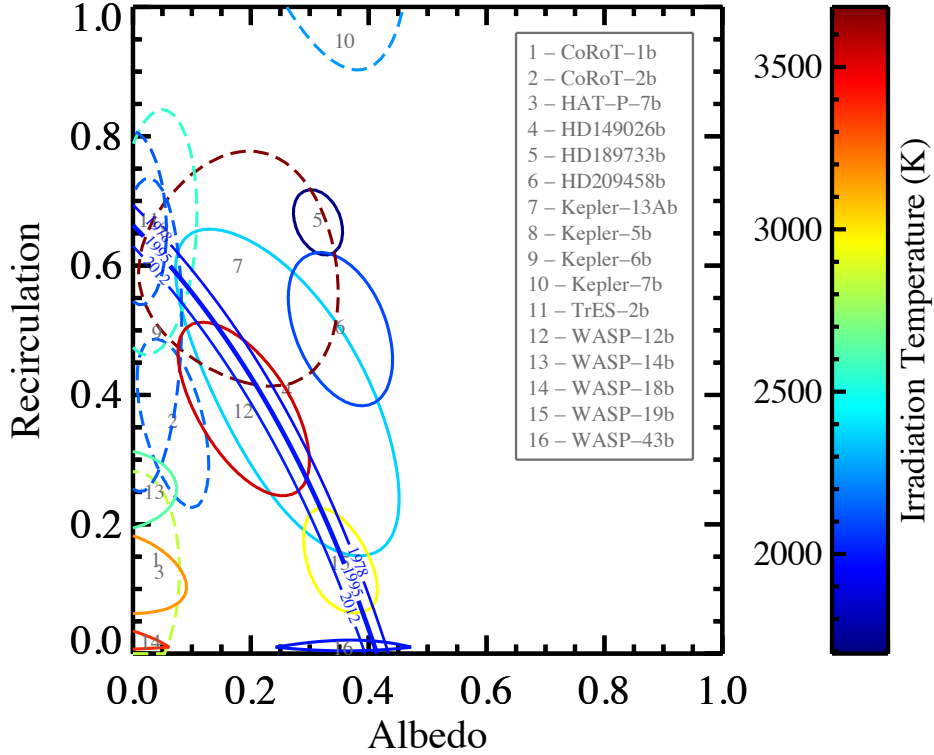}
\caption{Constraints on the planetary albedo and recirculation efficiency of HAT-P-32Ab  (blue curve with 1-$\sigma$ confidence regions) compared to measurements for other sixteen gas-giants (closed regions). The main reason for the degeneracy strip instead of a constrained 'island' for HAT-P-32Ab is in the limited constraining power of infrared eclipse observations alone. The degeneracy has been resolved for the sixteen exoplanets by combining infrared with optical occultations. Composite $1\sigma$ confidence regions are shown with the horizontal axis measuring different quantities: {\textit{Geometric albedo}} at visible wavelengths for exoplanets with optical eclipse observations-only (dashed lines) and {\textit{Bond albedo}} for thermal observation planets (solid lines). The colour bar indicates irradiation temperature, where the red and purple colours correspond to warmer and cooler temperatures, respectively.}
\label{fig:fig7}
\end{figure*}

The measured brightness temperature $(T_{{\rm{p}}})$ and 
consistency with a blackbody spectrum across a wide wavelength range allows us to 
explore the planet's bond albedo $A_B$ 
and redistribution efficiency $\varepsilon$, $0<\varepsilon<1$ using 
the \cite{cowan11} day-side temperature, $T_{\rm{d}}$ parameterization:
\begin{equation}
          T_{\rm{d}}=T_0(1-A_B)^{1/4} \bigg(\frac{2}{3}-\frac{5}{12}\varepsilon\bigg)^{1/4},
	\label{eq:td}
\end{equation}
where $T_0=T_{{\rm{eff}}}/\sqrt{a/R_{\ast}}$ is the equilibrium temperature at the planet's substellar point (for a circular orbit)
and $T_{{\rm{eff}}}$ is the effective temperature of HAT-P-32A.
Assuming that the day-side temperature, $T_{{\rm{d}}}$ is equal to the 
planet brightness temperature, $T_{{\rm{p}}}$ we solved 
Equation\,\ref{eq:td} for $A_B$ and propagated the uncertainty of $T_{{\rm{p}}}$ (Figure\,\ref{fig:fig7}). 
Our result rules out a low-albedo low-recirculation scenario for the atmosphere of HAT-P-32Ab. 
In addition, Figure\,\ref{fig:fig7} also implies that if the the albedo is low the planet 
would have an efficient recirculation.  

Previous findings \citep{cowan11, cowan12, perez13} have shown a 
tendency towards lower recirculation efficiency and greater day-night 
temperature contrasts as the stellar irradiation increases. Interestingly for 
HAT-P-32Ab, a high recirculation of 0.65 is suggested when 
assuming zero albedo, which given it's high $\sim1995$\,K 
day-side temperature would go against this trend.  
In a comprehensive study of fifty short-period transiting giant exoplanets 
\cite{schwartz15, schwartz17} resolved the albedo versus heat-transport degeneracy 
for sixteen exoplanets from 
the sample by considering eclipse measurements 
at optical wavelengths and phase curve data. 
The authors found evidence for reflective clouds and 
optical absorbers for some planets 
like HD189733b while others had very low albedos.
Comparing our Figure\,\ref{fig:fig7} to their Figure 7, if HAT-P-32Ab's 
albedo were low ($<0.15$) the 
recirculation would have to be more efficient ($\gtrsim0.6$) than most of the 
constrained exoplanets, whose recirculation range between about 0.4 to 0.7 
in the low albedo regime. 
Given the transmission spectrum of the planet indicates the 
presence of thick clouds \citep{gibson13a, Mallonn16a, Mallonn16b, Nortmann16} 
and the day-side temperature is close to several condensation curves, 
a high albedo for the planet seems plausible, and could be verified with 
optical eclipse or phase curve information.
Kepler-7b, another low-density bloated exoplanet,
has also been shown to have a high albedo in the range $0.4-0.5$, over the {\it{Kepler}} passband, \cite{garcia15}.  The authors used optical phase curve 
observations, which enabled constrains on the composition and cloud particle sizes, consistent
with condensates of silicates, perovskite and silica of sub-micron radii.

\subsection{Interpretation}
The combined WFC3, WIRC and IRAC day-side thermal emission spectrum of HAT-P-32Ab 
is compatible with an isothermal blackbody spectrum
at a temperature of\,\,$T_{\rm{p}}=1995\pm17$K (BIC$=30.1$), but can equally-well be described by 
a model, assuming thermal inversion (BIC$=28.6$\footnote{Bayesian Information Criterion, \cite{Schwarz78}}).
The WFC3 observation also shows no evidence of the absorption or emission signature of water at 
$1.4\mu$m (Figure\,\ref{fig:fig4} and \ref{fig:fig9}). Because the two models 
can be alternative explanations of our observations, 
and are both nearly equally likely statistically, 
we discuss the possible scenarios for the atmosphere of the planet 
assuming each of them.

\subsubsection{Isothermal spectrum}
In the case of an isothermal spectrum there are 
two distinct scenarios for the atmosphere of the planet.
In the first scenario, the spectrum probes 
pressure levels with a constant temperature, 
equal to the measured 
blackbody temperature. 
A constant temperature with altitude can be maintained 
if a cooling and heating mechanisms balance 
in the probed atmospheric layers.
For a clear atmosphere with a decreasing 
temperature profile, this can be the case 
of an absorber that traps stellar radiation
and effectively modifies 
the profile to resemble an isothermal or  a profile with weak 
inversion (i.e. increasing temperature with altitude).
Prime candidates for the hypothetical absorber could be 
the spectroscopically active TiO and VO, expected to be in a gas phase 
at the day-side temperature of the planet \citep{fortney08, lodders99}. 
Such balancing conditions are known to occur in the 
transition atmospheric layers, e.g. tropopause, stratopause, mesopause, etc. of 
solar system planets with rather limited depth of the order of a few tens kilometers.
If our combined emission spectrum is produced by such a hypothetical 
transition layer its depth would be substantial, ranging from $\sim3$bar to $\sim3$mbar.


In the second scenario, the observed isothermal spectrum 
could probe one and the same pressure level. In this case, the 
spectrum could be produced by a thick dusty cloud deck covering the day-side, 
with a temperature on its top equal to the measured blackbody temperature.
Near-infrared secondary eclipse observations 
can probe layers deep in the day-side of the atmosphere reaching a few bars.
However, in the case of a high-altitude cloud deck 
distinguishing between a high-altitude versus 
low-altitude cloud deck could virtually be 
impossible should the grain sizes are comparable or larger 
than the wavelength (e.g. microns). 
In the case of a low-altitude cloud deck, 
the layers above it could contribute with absorption or emission lines, 
should the layer temperature be lower or higher than the temperature 
of the deck, respectively. Our comparative 
analysis with models showed that absorption lines would 
easily be detectable, but are not present. There is still the possibility 
of hotter layers, contributing with emission lines, above the 
cloud deck, but their detection would be challenging at the 
precision of the data.
Hence we cannot distinguish 
between low-altitude and high-altitude cloud deck.

A cloudy day-side scenario would also agree with the evidence for clouds at the 
averaged day-night terminator.
Featureless flat optical transmission spectroscopy from 0.3 to $1\mu$m has been 
reported by a number of studies e.g. \cite{gibson13a, Mallonn16a, Mallonn16b, Nortmann16}
along with a recent detection of nearly half reduced amplitude water feature at $1.4\mu$m from 
{\it{HST}} WFC3 transmission spectroscopy \citep{tsiaras17}. 
Those observations imply a high-altitude dust deck or a layer of haze 
with grain sizes comparable to the wavelength of the the visible light. 
Due to the larger wavelength in the near-infrared the radiation starts 
to penetrate to deeper layers, which can explain the reduced water feature.  
Given the equilibrium temperature of the planet $1786\pm26$K, 
potential condensate forming species at the cooler day-side terminator  
include the silicate condensates enstatite and forsterite \citep{lodders99}.  
However, it should be noted that clouds at the 
terminator do not necessarily imply clouds on the day-side due to the 
significant differences in the temperature and altitude.

A third possible scenario to produce the observed blackbody spectrum 
would require a combination of 
partially clear and cloudy dayside atmosphere. 
This scenario is plausible given the fact that secondary eclipse 
observations probe also the region of the atmosphere around the planet's hot-spot, 
which has been predicted to be free of clouds, due to its high temperature and 
strong winds \citep{parmentier16}.

\subsubsection{An atmosphere with thermal inversion}
A model atmosphere with thermal inversion is an alternative 
explanation for the observed emission spectrum. 
An optical absorber could produce significant heating 
in the upper atmosphere to either cause 
an inversion or be near-isothermal over a large pressure range. 
VO is favoured by the data at vertical mixing ratio (VMR) of 
$2.3\times10^{-7}$ ($30\times$solar) and matches the WFC3 features, 
though the lower abundance range is unconstrained.
Presence of VO would be consistent with inverted scenario in 
which the day-side atmosphere is cloud-free. 
The signature of VO instead of water can be explained 
by the fact that the cross-section of VO is 
several orders of magnitude 
larger than the cross section of water at the WFC3 
wavelengths (\citealt{evans17}).



\subsubsection{Future observations}
Since planetary occultations only offer information of the dayside, they cannot resolve the circulation-Bond albedo degeneracy. Optical secondary eclipse or phase curve observations or alternatively infrared phase curves can help resolve this degeneracy. In addition, observations at higher signal to noise ratio and complementary wavelength regions can further shed light on the atmosphere of the planet. In particular observations with the {\it{James-Webb Space Telescope}} ({\it{JWST}}) can help resolve the three case scenario by detecting the signature of clouds at mid-infrared wavelengths. WFC3 hints at low-amplitude molecular features in the spectrum which could be resolved. The thermal emission spectrum of HAT-P-32Ab joins a significant sample of isothermal spectra reported in the literature (e.g. \citealt{cartier17, crouzet14, stevenson14c}).



\section{Conclusions}\label{conclusions}
We reported {\it{HST}} WFC3
emission spectrum of HAT-P-32Ab
covering the wavelength range from 1.123 to 1.644\,$\mu$m.
Combined with previous thermal eclipse observations, our spectrum 
can equally-well be described by an isothermal blackbody spectrum with 
a temperature of $T_{{\rm{p}}}=1995\pm17$\,K or a model assuming 
thermal inversion layer.
A comparative and retrieval analysis with 1D radiative-convective atmospheric models excludes models, assuming 
non-inverted temperature profiles at a high confidence.
A blackbody or thermally inverted emission spectrum can 
imply several alternative scenarios for the 
planetary atmosphere, including i) clear atmosphere with absorber; ii) dusty cloud deck, 
or ii) combination of both. Eclipse observations with the JWST could help to potentially resolve
the three-case degeneracy for the day-side atmosphere of the planet. 
Converting the measured blackbody temperature 
to brightness temperature, we find that the planet can have continuum of values for the albedo and recirculation, ranging from high albedo and poor recirculation (bright and still) to low albedo but very efficient recirculation (dark and windy).
Optical eclipse observations 
could help to resolve the albedo versus recirculation degeneracy 
and shed more light on the presence or absence of clouds on the 
day-side of the planet. 

\begin{figure}
\includegraphics[trim = 0 0 0 0, clip, width = 0.5\textwidth]{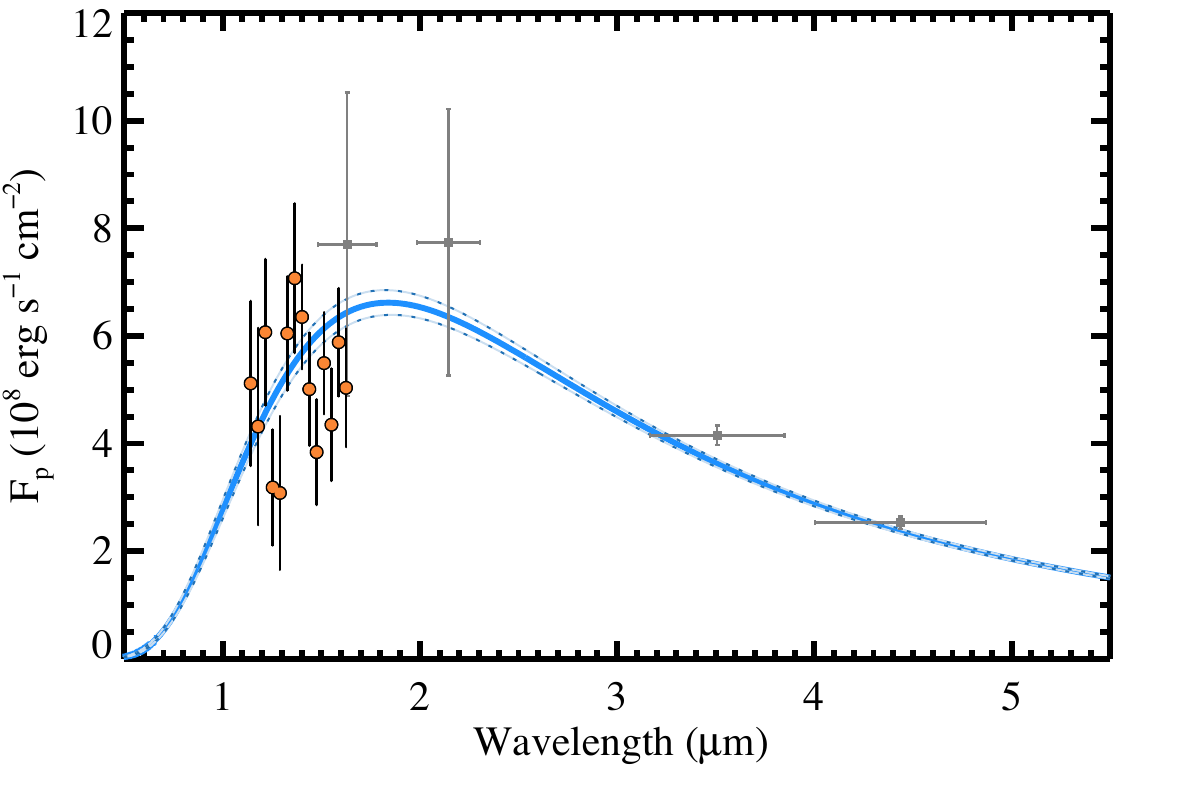}
\caption{Planet thermal emission spectrum of HAT-P-32Ab (orange and grey dots refer to the WFC3 and WIRC+IRAC observations, respectively) along with the best-fit blackbody curve ($T_{\rm{p}}=1995\pm17$\,K) and $1\sigma$ uncertainty (blue lines).}
\label{fig:fig8}
\end{figure}

\section*{Acknowledgements}
The authors are happy to thank D. Foreman-Mackey, J. Schwartz  and V. Parmentier
for fruitful discussions and help during the course of this work. 
This work is based on observations with the NASA/ESA {\it{Hubble Space Telescope}}, obtained at the Space Telescope Science Institute (STScI) operated by AURA, Inc. This work is based in part on observations made with the {\it{Spitzer Space Telescope}}, which is operated by the Jet Propulsion Laboratory, California Institute of Technology under a contract with NASA. N. N, D. K. S, and T. M. E. acknowledge funding from the European Research Council under the European Unions Seventh Framework Programme (FP7/2007-2013) / ERC grant agreement no. 336792. J. G. acknowledges support from a Leverhulme Trust Research Project Grant. G.W.H. and M.H.W. acknowledge long-term support from Tennessee State University and the State of Tennessee through its Centers of Excellence program and from the Space Telescope Science Institue under HST-GO-14767. The authors would like to acknowledge the anonymous referee for their useful comments.









\appendix
\section{White light curve MCMC posterior distributions}\label{appendix}
\begin{figure*}
\includegraphics[trim = 0 0 0 0, clip, width = 0.99\textwidth]{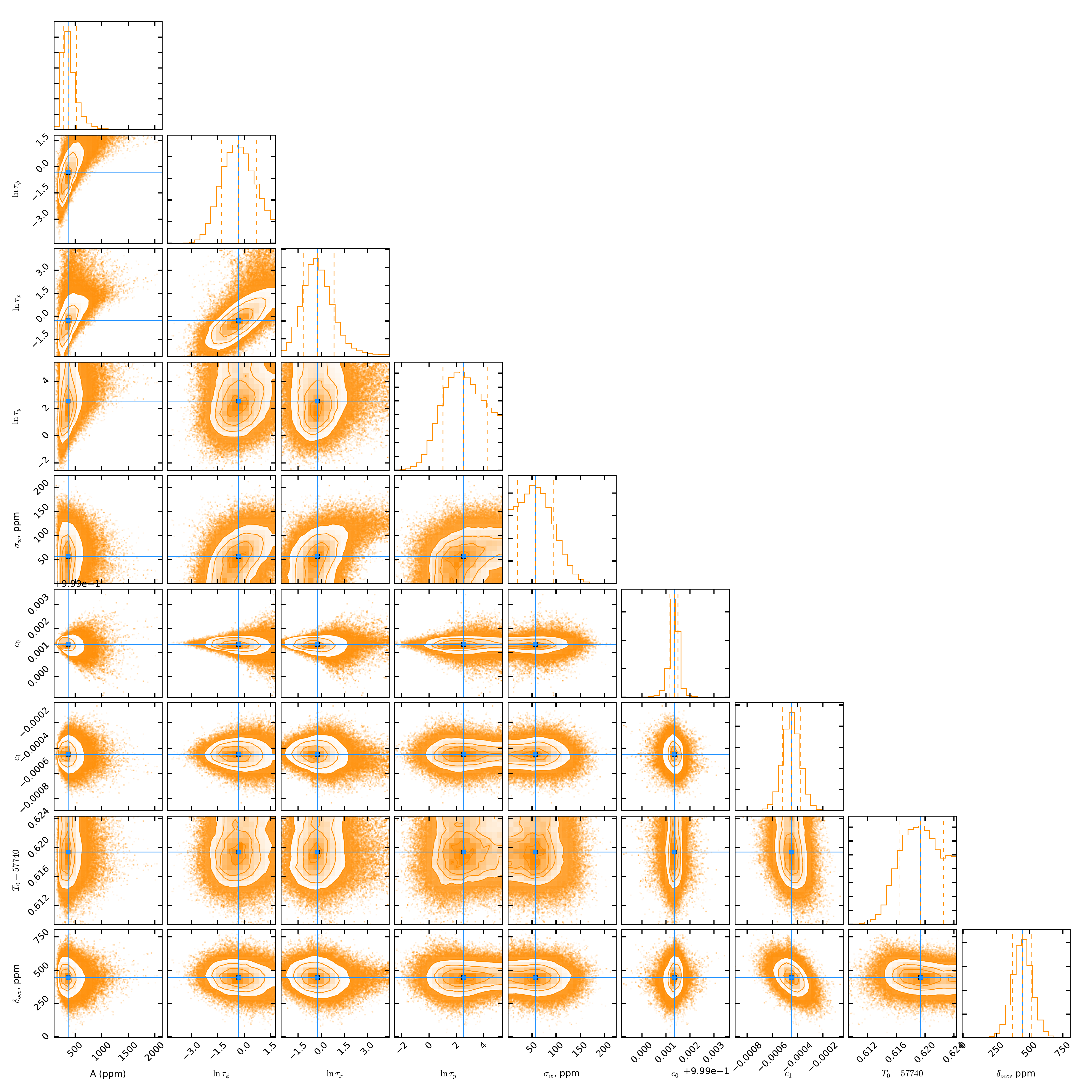}
\caption{Posterior distributions of the variable eclipse parameters and hyperparameters from the MCMC chains for the white light curve. The scatter plots show all pairs of parameters plotted after marginalization over all other parameters, and the histograms show the marginalized posterior distribution of each individual parameter. The median and 1-$\sigma$ measured parameters are indicated with continuous and dashed lines, respectively.}
\label{fig:fig6}
\end{figure*}

In this appendix we present posterior distributions of the variable eclipse parameters and 
hyperparameters from the MCMC chains for the white light curve of HAT-P-32Ab (Figure\,\ref{fig:fig6}).

\section{Spectroscopic light curves}\label{appendix2}
\begin{figure*}
\includegraphics[trim = 0 0 0 0, clip, width = 0.98\textwidth]{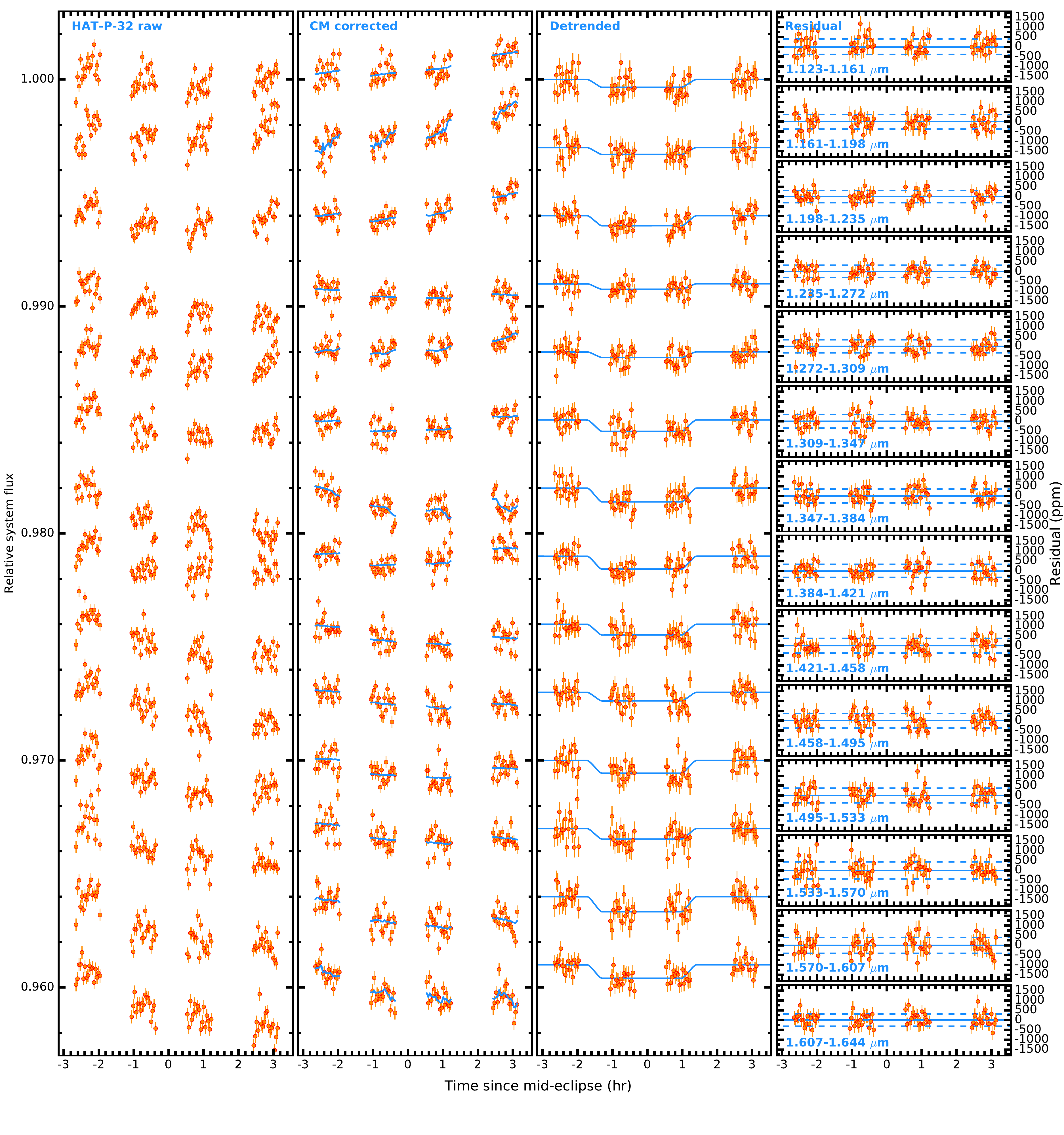}
\caption{Spectroscopic light curves for HAT-P-32Ab. {\it{First panel}}: Raw normalised light curves.
{\it{Second panel}}: Common-mode corrected raw light curves along with the best-fit GP systematics and eclipse model.
{\it{Third panel}}: Detrended eclipse light curves along with an eclipse model calculated from the marginalized posterior distributions.
{\it{Small panels}}: Light curve residuals with error bars from each fit.}
\label{fig:fig3}
\end{figure*}

\section{Retrieval results}\label{appendix3}
\begin{figure*}
\includegraphics[trim = 0 0 0 0, clip, width = 0.98\textwidth]{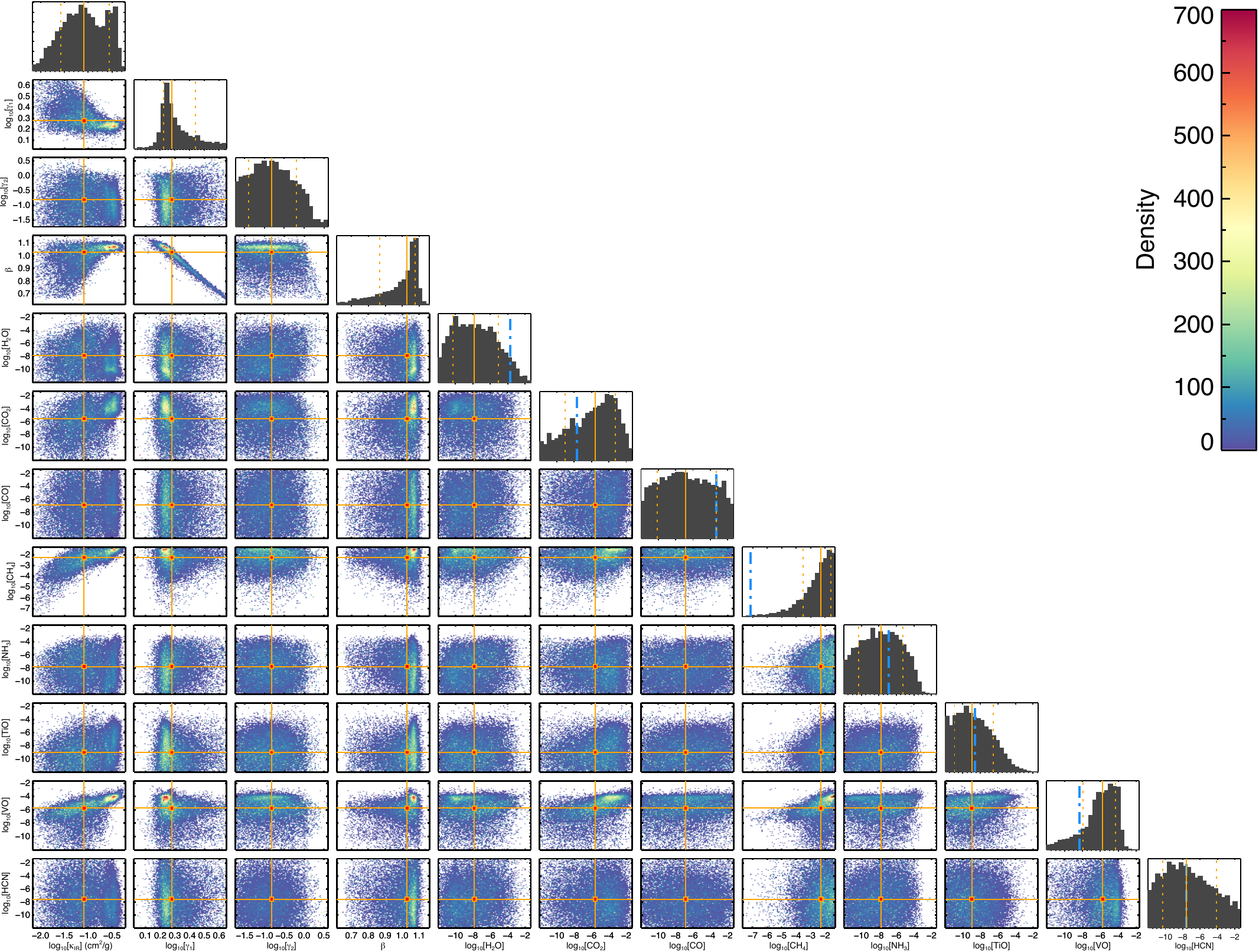}
\caption{ Posterior distributions from the MCMC retrieval analysis. 
The scatter plots show all pairs of parameters plotted after marginalization over all other parameters, and the histograms show the marginalized posterior distribution of each individual parameter. The median and 1-$\sigma$ measured parameters are indicated with continuous and dashed lines, respectively. The dash-dotted lines represent the solar abundances, calculated with {\tt{ATMO}}, assuming solar metallicity and solar C/O ratio and a recirculation factor of 0.75 at 1 bar with rainout. }
\label{fig:fig10}
\end{figure*}



\bibliographystyle{mnras}
\bibliography{researchv3}

\end{document}